# DETERMINATION OF THE SYMMETRIES OF AN EXPERIMENTALLY DETERMINED STIFFNESS TENSOR; APPLICATION TO ACOUSTIC MEASUREMENTS

abbreviated title: "symmetries of a stiffness tensor experimentally obtained"

M. François, G. Geymonat and Y. Berthaud*
Laboratoire de Mécanique et Technologie, URA 860 ENS de Cachan-CNRS-Université Paris 6, 61 avenue du président Wilson, 94235 Cachan Cedex.
*Laboratoire des Matériaux et des Structures du Génie Civil, UMR 113 LCPC-CNRS,
2 allée Kepler, 77420 Champs-sur-Marne.
e-mail: francois@lmt.ens-cachan.fr, geymonat@lmt.ens-cachan.fr, berthaud@umrserv.inrets.fr

## Abstract

For most materials, the symmetry group is known *a priori* and deduced from the realization process. This allows many simplifications for the measurements of the stiffness tensor. We deal here with the case where the symmetry is *a priori* unknown, as for biological or geological materials, or when the process makes the material symmetry axis uncertain (some composites, monocrystals). The measurements are then more complicated and the raw stiffness tensor obtained does not exhibit any symmetry in the Voigt's matricial form, as it is expressed in the arbitrarily chosen specimen's base. A complete ultrasonic measurement of the stiffness tensor from redundant measurements is proposed. In a second time, we show how to make a plane symmetry pole figure able to give visual information about the quasi-symmetries of a raw stiffness tensor determined by any measurement method. Finally we introduce the concept of distance from a raw stiffness tensor to one of the eight symmetry classes available for a stiffness tensor. The method provides the nearest tensor (to the raw stiffness tensor) possessing a chosen symmetry class, with its associated natural symmetry base.

## 1. Introduction

### 1.1. Basic problem: the material symmetries are unknown

This paper addresses a general problem: the knowledge of the symmetry class and the symmetry axis of a material whose stiffness tensor is obtained through different experimental techniques, either mechanical or acoustic. For most of the materials, the manufacturing process implies the material symmetry. However, the symmetry axis can be hardly known when the process is difficult to control: the growing direction of our monocrystal (made of the γ phase of a nickel-based superalloy) is not precisely known, although the atomic disposition involves a cubic symmetry. In the case of geological materials the symmetry class is often unknown. No simplification is then available on the measurement of the stiffness tensor, the 21 constants have to be identified. This tensor has to be studied to point out the possible symmetry class of the material.

### 1.2. Measurement of the entire stiffness tensor

The first step is to determine the stiffness tensor of the material before analyzing the possible symmetries of this tensor. Hayes(1969) has solved the theoretical problem of the determination of the stiffness tensor **C** even if the material symmetry is unknown. The proposed mechanical tests appear to be difficult to perform. Six classic tensile tests in six independent directions (François, 1995) are also able to give **C** once the strains in different directions are measured for each test. This later method implies a tedious machining and off-axis mechanical tests that are always difficult to perform (Boehler *et al.*, 1994). In the particular case of a priori known symmetries the identification of the symmetric tensor **Cs** is easily done (up to orthotropy) with various tensile specimens machined with respect to the symmetries of the material.

The ultrasonic measurements are used for biological (Van Bursirk *et al.*, 1986) or geological (Harder, 1985) materials because they are the only method that allows the measurement of the elastic properties on very small specimens. The transducers are directly stuck on parallel faces of the sample. This is called the direct contact method. The direction of the propagation of the ultrasonic waves is then fixed by the geometry of the specimen. To avoid the problem of coupling between the transducer and the material and to examine various directions of propagation, the immersion method, in which transducers are sending the waves to the sample through immobile water, is preferred for plate-shaped composite materials (Castagnède *et al.*, 1990) Combined with digital signal processing, this technique ensures a high accuracy of the measurement of the velocity and involves continuously varying directions of the wave propagation as the plate is rotated. This method has been fully developed and is practically able to give stiffness tensors up to orthotropy either in the case of exactly aligned specimen or non aligned specimen (Baste and Hosten, 1990; Chu *et al.*, 1994) but does not seem to have yet been applied to a real triclinic material. The direct contact method has been retained in our experiments on small specimens. This technique allows the determination of the raw stiffness tensor **Co** in the base Bo linked to the specimen. The expression *raw stiffness tensor* refers to a tensor that is perturbed by errors due to uncertainties of the measurements, to a possible non homogeneity of the tested specimen.

### 1.3. Quasi symmetries: indicators and visualization

At this step we suppose a raw stiffness tensor **Co** determined by some experimental method in the specimen's base Bo, for example the acoustical method. The 6x6 Voigt's representation table of **Co** does not exhibit generally the symmetry of the stiffness tensor as the base Bo has no relation with the (unknown) possible principal directions of the material. Furthermore the raw tensor **Co** is perturbed by experimental errors. It has generally no exact symmetry (triclinic) but may be close to a symetry level (in this case we say that it has a *quasi symmetry* ). This forbids (or makes complex) the use of classical indicators like the contractions $C_{ijkk}$ and $C_{ikkj}$ (Jaric, 1994) of a stiffness tensor **C**.

In order to have a qualitative information on the quasi symmetries of **Co**, we developed a map based on the crystallographic pole figures. These maps represent, on a colored scale, the relative discrepancy between **Co** and its symmetric according to the plane of normal **r**. The "spots" or "lines" indicate the normals of the planes for which **Co** is close to be monoclinic. Their number and their relative position reveal the level(s) of symmetry(ies) **Co** is close to. For example, we will see that the monocrystal exhibits clearly nine spots corresponding to the cubic symmetry although the tensor is *stricto sensu* triclinic.

### 1.4. The nearest symmetric stiffness tensor

Once a symmetry class is chosen using the above pole figure for the stiffness tensor we have to compute the symmetric tensor **Cs** belonging to the chosen class of symmetry and that is the nearest to the raw tensor **Co**. The natural base Bs is the base for which the tensor **Cs** has the classical form in the 6x6 Voigt matricial representation. One way to find **Cs** and Bs is to look for the base B that minimizes a function deduced from the classical relations between the components of the tensor that allow the Voigt's representation (Arts, 1993). The choice of such minimization functions remains arbitrary and leads to non-intrinsic functions.

We propose an intrinsic function that creates from **Co** and the arbitrary base B a tensor **Cb** which has the chosen symmetry group G. This function calculates the average of **Co** on its orbit according to $G_B$ related to the base B. The natural symmetry base Bs is the one for which the *relative discrepancy* D(B) between **Cb** and **Co** is minimum. Then **Cb** is **Cs**, the nearest (to **Co**) symmetric stiffness tensor and D(Bs) can be called the distance from **Co** to **Cs**; in other words the distance to the symmetry group G.

This knowledge can be useful in other cases than the basic problem described above. We can imagine, for example, the simplified resolution of some mechanical problems while using a tensor of a higher symmetry group (fewer independent components), with some acceptable error.

## 2. Acoustic measurements

### 2.1. Description of a direct contact measurement

The propagation of acoustic waves in solid media is mainly described by the acoustic or Cristoffel's tensor $\Gamma$ (Auld, 1973). It is linked to the unknown stiffness tensor **C** and for a given direction **n** of the wave propagation by the following relation:

$$\Gamma(\mathbf{n}) = \mathbf{n}.\mathbf{C}.\mathbf{n} \qquad (1)$$

This symmetric second order tensor $\Gamma(\mathbf{n})$ can be written in a diagonal form. Each eigenvector (at least three) represents a direction of vibration of the particles $\mathbf{u}^i$, and each corresponding eigenvalue represents the product $\rho(V^i)^2$ where $\rho$ is the mass density, and $V^i$ the velocity of the body wave polarized in the direction $\mathbf{u}^i$. The direction $\mathbf{u}^1$, the closest to the propagation direction **n**, is called the *quasi-longitudinal* wave and the two others, $\mathbf{u}^{2,3}$, the *quasi-transverse* waves. This allows us to write $\Gamma(\mathbf{n})$ in the following different form (where "$\otimes$" represents the tensorial product):

$$\Gamma(\mathbf{n}) = \sum_{i=1}^{3} \rho(V^i)^2 \, \mathbf{u}^i \otimes \mathbf{u}^i \qquad (2a)$$

Let us call **m** the *direction of the vibration* (polarization vector) of both the ultrasonic transducers: the emitter and the receiver. They are stuck on two parallel faces of the specimen. In the case of *transverse* emission **m** is in the plane orthogonal to **n** and the transducers are coupled by a rod (Figure 1). In the case of *longitudinal* emission **m** is equal to **n**. The generated wave polarized in the direction **m** is decomposed in the material in the three directions $\mathbf{u}^i$ (with a displacement proportional to the scalar product $\mathbf{u}^i.\mathbf{m}$) allowed by the material and each wave propagates at the velocity $V^i$. As these waves reach successively the receiver the displacement measured by the receiver is given by $\mathbf{u}^i.\mathbf{m}$. The measure (supposing that the transducer has a linear response) is then proportional to $(\mathbf{u}^i.\mathbf{m})^2$.

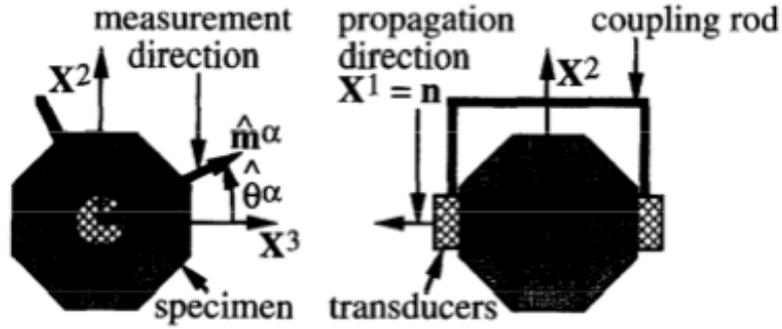

Figure 1: experimental set-up for acoustical measurements.

This allows the operator, when using transverse transducers (**m.n**=0), to measure the *polarization angle* $\hat{\theta}^i$ from the direction $X^3$ to the direction $X^2$ in the set-up coordinates $X^i$ ($X^1$ is equal to the direction of propagation **n**) (Figure 1), while searching for the maximum of received signal, *i.e.* when $u^i.m$ is maximum (Figure 2). Pratically this is not possible for the quasi-longitudinal wave as $u^1.n$ is close to 1, $u^1.m$ is close to 0 and too weak to be measured: $\hat{\theta}^1$ cannot be measured with our equipment.

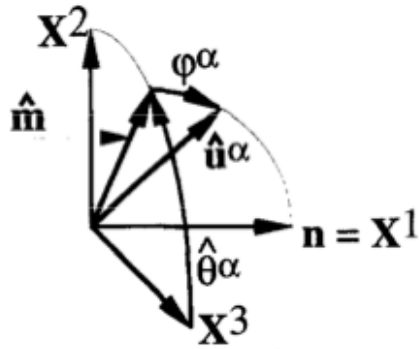

Figure 2: measurement of the polarization $\hat{\theta}^\alpha$ with transversal transducers.

For each direction **n**, the three wave speeds $\hat{V}^i$ and the two polarization $\hat{\theta}^\alpha$ angles (with $\alpha = 2, 3$) of the quasi-transverse waves are finally measured. In this ideal case the measure is called a *complete measure*. The two deflection angles $\varphi^\alpha$ between the plane $n^\perp$ (orthogonal to **n**) and $u^\alpha$ (Figure 2) remain unknown. For each direction **n** we can define the measured acoustic tensor $\hat{\Gamma}(\mathbf{n})$ equal to:

$$\hat{\Gamma}(\mathbf{n}) = \sum_{i=1}^{3} \hat{\rho}(\hat{V}^i)^2 \, \hat{\mathbf{u}}^i \otimes \hat{\mathbf{u}}^i \quad (2b)$$

All the measured values of a quantity x are denoted by $\hat{x}$. The tensor $\hat{\Gamma}$ should be equal to $\Gamma$ (eqn 1) if the measured values $\hat{\rho}$, $\hat{V}^i$ and $\hat{\mathbf{u}}^i$ were exact. For each measurement along the direction **n** we have five experimental values ($\hat{V}^i$ and $\hat{\theta}^\alpha$) and two additional unknowns ($\varphi^\alpha$). Then the determination of the 21 values of the unknown stiffness tensor requires at least 7 different directions **n** (in case of complete measurements).

## 2.2. The specimen and the measures
**equipment**

The geometry of the specimen is a compromise between simple machining, high number of faces and small specimen. It has here 13 pairs of parallel faces (Figure 3). This will give redundant data (we have seen that the minimum is 7). This redundancy will reduce the effect of the experimental errors, through a minimization scheme. The equidistant faces are cut orthogonal to the three vectors $\mathbf{x}^{o1}$, $\mathbf{x}^{o2}$, $\mathbf{x}^{o3}$ of the *specimen's base* $\mathcal{B}o$, to its bisectors, and the same after a rotation along $\mathbf{x}^{o2}$ (Table 1). We used 5 MHz "Panametrics" transducers with a 12 mm diameter and a "Saphir" card integrated in a PC computer. Software has been developed to measure the velocities (using an intercorrelation technique) and the directions $\hat{\theta}^{\alpha}$.

| face | A | B | C | I | J | K | L | M | N | $\alpha$ | $\beta$ | $\gamma$ | $\delta$ |
|---|---|---|---|---|---|---|---|---|---|---|---|---|---|
| normal's | 1 | 0 | 0 | 0 | $1/\sqrt{2}$ | $1/\sqrt{2}$ | 0 | $-1/\sqrt{2}$ | $-1/\sqrt{2}$ | 1/2 | -1/2 | -1/2 | 1/2 |
| coordinates | 0 | 1 | 0 | $1/\sqrt{2}$ | 0 | $1/\sqrt{2}$ | $1/\sqrt{2}$ | 0 | $1/\sqrt{2}$ | $1/\sqrt{2}$ | $1/\sqrt{2}$ | $1/\sqrt{2}$ | $1/\sqrt{2}$ |
| in $\mathcal{B}o$ | 0 | 0 | 1 | $1/\sqrt{2}$ | $1/\sqrt{2}$ | 0 | $-1/\sqrt{2}$ | $1/\sqrt{2}$ | 0 | 1/2 | 1/2 | -1/2 | -1/2 |
| $\mathbf{X}^2$ | B | C | A | A-L | -B | -C-N | -A-I | -B | C-K | -M-$\gamma$ | J-$\delta$ | M-$\alpha$ | -J-$\beta$ |

Table 1: position of the specimen's faces.

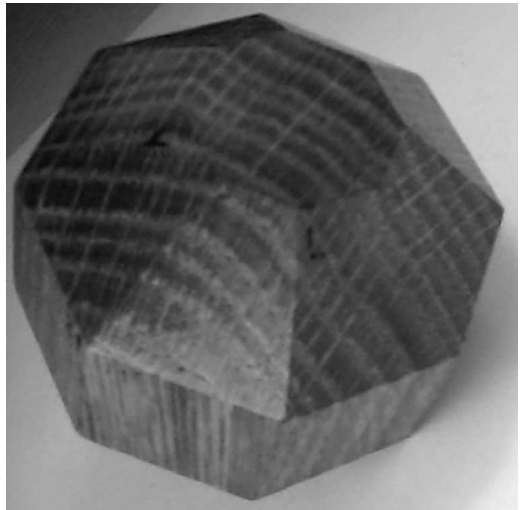

Figure 3: 26-faces oak specimen.

**practical experimentation**

Table 2 gives the values of the velocities $V^i$ and of the angles $\hat{\theta}^{\alpha}$ for all the faces of the superalloy specimen. The distance between two parallel faces is 18.5 mm. The data given (Table 2) have been obtained after three measurements (under the same conditions) on the same specimen. This allows us to evaluate the uncertainty of the experimental procedure. The accuracy of the velocity measurement can be estimated to ± 3% and to be about ± 5° for the angle. We can remark that the values given for the faces A, C, $\alpha$ and $\delta$ are uncertain because the signal was weak and distorted. They will be removed from the set of data. As a consequence it is possible to have either complete measurement (faces B, I , J, K, L, M and N for example) or incomplete measurement if the other faces are also considered to get more experimental data. The following two sections will deal with the case of complete and incomplete measurements.

| face | A | B | C | I | J | K | L | M | N | α | β | γ | δ |
|---|---|---|---|---|---|---|---|---|---|---|---|---|---|
| $\hat{V}^1$ (m/s) | 5437 | 5316 | 6131 | 6098 | 6066 | 6098 | 6151 | 6066 | 6165 | 6275 | 6244 | 6296 | 6307 |
| $\hat{V}^2$ (m/s) | 3826 | 3863 | 3867 | 3918 | 3826 | 3840 | 3959 | 3863 | 3955 | 3615* | 3615* | 3676 | 2712 |
| $\hat{\theta}^2$ | 0 | 0 | 115 | 115 | 165 | 110 | 125 | 175 | 135 | 130 | 130 | 145 | 37* |
| $\hat{V}^3$ (m/s) | 1869* | 3863 | 1815* | 2237 | 2143 | 2212 | 2134 | 2154 | 2244 | 2674 | 2396 | 2460* | 3615 |
| $\hat{\theta}^3$ | 90 | 90 | 25* | 25 | 69 | 35 | 35 | 85 | 37 | 45 | 35* | 55* | 145* |

Table 2 : acoustical measurements for the monocrystal. The * indicates doubtful data.

## 2.3. Discrepancy tensor and minimization function J

The basic idea is to find the set of components of **Co** that makes every acoustic tensor **Γ(n)** (eqn 1) as close as possible to every experimental acoustic tensor $\hat{\Gamma}$(**n**) (eqn 2b). For every measure (for every **n**) we define the function j(**n**) from the Euclidean norm of the discrepancy between **Γ(n)** and $\hat{\Gamma}$(**n**) :

$$j(\mathbf{n}) = \frac{1}{2} (\Gamma(\mathbf{n}) - \hat{\Gamma}(\mathbf{n}))^2 \qquad (3)$$

**minimization strategy**

The "best" raw tensor **Co** (of every **C**) and the best deflection angles $\varphi^\alpha$ are those that minimize the sum J of j over the set of different directions **n** with complete measurements. As the two sets of variables (the components of **C** and the deflection angles $\varphi^\alpha$) have a completely different role, we chose to use an iterative process: the minimisation will be done alternately on each set. The first step is relative to the components of **C**, while the deflection angles $\varphi^\alpha$ are set at zero at the beginning of the iterative procedure (this is the value corresponding to an isotropic tensor).

**minimization of J with respect to the components $C_{ijkl}$**

The condition for J to be a minimum with respect to the components of **C** is the nullity of the following gradient:

$$\frac{\partial j(\mathbf{n})}{\partial \mathbf{C}} = \mathbf{n} \otimes (\mathbf{n}.\mathbf{C}.\mathbf{n}) \otimes \mathbf{n} - \sum_{i=1}^{3} \hat{\rho}(\hat{V}^i)^2 \, \mathbf{n} \square \, \hat{\mathbf{u}}^i \square \, \hat{\mathbf{u}}^i \square \, \mathbf{n} \qquad (4)$$

The measured $\hat{\theta}^\alpha$ and the unknown $\varphi^\alpha$ define $\hat{\mathbf{u}}^\alpha$ but not $\hat{\mathbf{u}}^1$. This vector $\hat{\mathbf{u}}^1$ is assumed to be orthogonal to $\hat{\mathbf{u}}^2$ and $\hat{\mathbf{u}}^3$ as required for exact measurements[1]. The classic symmetries of **C** are guaranteed by 60 Lagrangian multipliers $\mu_m$. We show the first of these terms

$$\mu_1 (C_{1112} - C_{1121}) = 0 \qquad (5)$$

The 81 equations (4) and the 60 equations (5) are disposed in a matrix form A.X = $\hat{B}$ in which the $\mu_m$ and the $C_{ijkl}$ are stored in the "vector" X, their multipliers in A and the second members of equations (4) in $\hat{B}$. This linear matricial system is directly and quickly resolved, giving the best components $C_{ijkl}$ of **C** for the given deflection angles $\varphi^\alpha$.

**minimization of each j with respect to the deflection angles $\varphi^\alpha$**

We now have to minimize J with respect to the deflection angles $\varphi^\alpha$, with **C** fixed. As the $\varphi^\alpha(\mathbf{n})$ play an independent role in each j(**n**), we can minimize independently each

---

[1] Even if $\hat{\mathbf{u}}^2$ and $\hat{\mathbf{u}}^3$ are not exactly orthogonal at this step of the calculation.

j(**n**) with respect to the two deflection angles. The gradient of j with respect to the deflection angles $\varphi^\alpha$ is given by the following equations (in which "∧" represents the vectorial product):

$$\frac{\partial j(\mathbf{n})}{\partial \varphi_\alpha} = - (\Gamma(\mathbf{n}) - \hat{\Gamma}(\mathbf{n})) : \sum_{i=1}^{3} \hat{\rho}(\hat{V}^i)^2 \left( \hat{\mathbf{u}}^i \Box \frac{\partial \hat{\mathbf{u}}^i}{\partial \varphi_\alpha} + \frac{\partial \hat{\mathbf{u}}^i}{\partial \varphi_\alpha} \Box \hat{\mathbf{u}}^i \right) \quad (6)$$

with 
$$\frac{\partial \hat{\mathbf{u}}^\alpha}{\partial \varphi_\beta} = \delta_{\alpha\beta} (- \sin\varphi_\alpha \mathbf{m}^\alpha + \cos\varphi_\alpha \mathbf{n}) \quad (7)$$

and 
$$\frac{\partial \hat{\mathbf{u}}^1}{\partial \varphi_\alpha} = \mathbf{v}^\alpha - \hat{\mathbf{u}}^1 (\mathbf{v}^\alpha . \hat{\mathbf{u}}^1) \quad (8)$$

with 
$$\mathbf{v}^\alpha = \frac{\left[ \frac{\partial (\hat{\mathbf{u}}^2 \Box \hat{\mathbf{u}}^3)}{\partial \varphi_\alpha} \right]}{\| \hat{\mathbf{u}}^2 \Box \hat{\mathbf{u}}^3 \|} \quad (9)$$

The knowledge of this gradient allows us to solve this non-linear problem through a B.F.G.S. (Burlish and Stoer, 1980) method.

**convergence - stopping criteria**

Two criteria are retained for stopping the calculation. A first test is done on the value of J which can be zero if the measurements are simulated ones, and the second involves the rate of convergence. This rate commonly tends to $10^{-2}$ after ten iterations.

**practical procedure**

One can compute the values of the speeds and of the polarizations given by this stiffness tensor **Co** and compare these values to the experimental ones. This may be useful to check doubtful measurements. The energy velocities are also computed to distinguish the waves that cannot reach the receiver without parasite reflection. This is the case for a strong anisotropy and for measurement direction **n** that are far from the symmetry axis. In such a case the corresponding measures are removed. Sometimes the received signal may be too weak to distinguish clearly a shear wave or to obtain its angle $\hat{\theta}^\alpha$. In these cases the set of measurements is incomplete.

## 2.4. Incomplete measurements

Let us suppose that the speed $\hat{V}^k$ (and of course, if it corresponds to a transverse wave, the direction $\hat{\mathbf{m}}^k$) has not been detected or corresponds to very doubtful data (see Table 2). This situation is more realistic especially with natural material (biology, geology).

We introduce the component $L_{ij}$ from the components of $(\Gamma - \hat{\Gamma})$ in the base $\hat{\mathbf{u}}^i \otimes \hat{\mathbf{u}}^j$ (i, j ≠ k).

$$L_{ij} = (\hat{\mathbf{u}}^i \otimes \mathbf{n}) : \mathbf{C} : (\mathbf{n} \otimes \hat{\mathbf{u}}^j) - \delta_{ij} \hat{\rho}(\hat{V}^i)^2 \quad (10)$$

We now define the new function h replacing j by:

$$h(\mathbf{n}) = \frac{1}{2} \sum_{i,j}^{\text{knowns}} L_{ij}^2 \quad (11)$$

This definition is similar to that of j (equation 3) if the $\hat{\mathbf{u}}^i$ are orthogonal. The global minimization function J has to be the sum of all the functions j(**n**) corresponding to complete measures and all the functions h(**n**) corresponding to incomplete ones. This theory cannot work for a measurement in which two speeds are not known. If the unknown polarization is quasi transverse, *e.g.* α = 3, a polarization angle $\hat{\theta}^3$ is missing and forbids us to calculate the quasi-longitudinal direction of vibration $\hat{\mathbf{u}}^1$ as before (orthogonal to $\hat{\mathbf{u}}^2$ and $\hat{\mathbf{u}}^3$). We know that the $\hat{\mathbf{u}}^i$ have to be an orthonormed base. This is written:

$$\sin\varphi_2 \sin\varphi_3 + \cos\varphi_2 \cos\varphi_3 \cos(\hat{\theta}^2 - \hat{\theta}^3) = 0 \qquad (12)$$

Given the variables $\varphi_2$ and $\varphi_3$, this condition gives two solutions for $\hat{\theta}^3$ (so for $\hat{\mathbf{u}}^3$); one of them is good (gives the lowest value for j). The gradient of $h(\mathbf{n})$ with respect to $\mathbf{C}$ is given by:

$$\frac{\partial h(\mathbf{n})}{\partial \mathbf{C}} = \sum_{i,j}^{known} L_{ij} \, \mathbf{n} \otimes \hat{\mathbf{u}}^i \otimes \hat{\mathbf{u}}^j \otimes \mathbf{n} \qquad (13)$$

Its structure is still linear with respect to the $C_{ijkl}$ and allows us to complete the matricial direct resolution described above. The gradient with respect to the variables $\varphi_\alpha$ allowing a B.F.G.S. resolution is given by

$$\frac{\partial h(\mathbf{n})}{\partial \varphi_\alpha} = \sum_{i,j}^{known} L_{ij} \left( \mathbf{C} :: (\mathbf{n} \otimes \frac{\partial \hat{\mathbf{u}}^i}{\partial \varphi_\alpha} \otimes \hat{\mathbf{u}}^j \otimes \mathbf{n} + \mathbf{n} \otimes \hat{\mathbf{u}}^i \otimes \frac{\partial \hat{\mathbf{u}}^j}{\partial \varphi_\alpha} \otimes \mathbf{n}) \right) \qquad (14)$$

We detail here the derivatives of $\hat{\mathbf{u}}^i$. If the unknown polarization corresponds to the quasi-longitudinal wave the calculation remains the same as before (eqn 7, 8, 9). If the unknown polarization corresponds to the quasi-transverse wave $\hat{\mathbf{u}}^3$, the derivative $\partial \hat{\mathbf{u}}^2/\partial\varphi^2$ is calculated as before (equation 7), but $\hat{\mathbf{u}}^1$ and its derivative now depend on the value of $\hat{\theta}^3$ itself dependent on $\varphi^3$ (equation 12). We have, in this case:

$$\frac{\partial \hat{\mathbf{u}}^1}{\partial \varphi_2} = -\sin\varphi_2 \, \hat{\mathbf{m}}^2 \wedge \hat{\mathbf{u}}^3 + \cos\varphi_2 \, \mathbf{n} \wedge \hat{\mathbf{u}}^3 \qquad (15)$$

$$\frac{\partial \hat{\mathbf{u}}^1}{\partial \varphi_3} = -\sin\varphi_3 \, \hat{\mathbf{u}}^2 \wedge \hat{\mathbf{m}}^3 + \cos\varphi_3 \, \hat{\mathbf{u}}^2 \wedge \mathbf{n} +$$

$$\pm \frac{\tan\varphi_2}{\cos\varphi_3} \frac{1}{\sqrt{1-\tan^2\varphi_2 \tan^2\varphi_3}} \left[ \cos\varphi_2 \cos\varphi_3 \cos(\hat{\theta}^3-\hat{\theta}^2) \, \mathbf{n} - \sin\varphi^2 \, \hat{\mathbf{m}}^3 \right] \qquad (16)$$

The value of $\pm$ is given by the choice of the best $\hat{\theta}^3$ described before (equation 12).

**practical results**

$$\begin{vmatrix} 243 & 136 & 135 & 22 & 52 & -17 \\ 136 & 239 & 137 & -28 & 11 & 16 \\ 135 & 137 & 233 & 29 & -49 & 3 \\ 22 & -28 & 29 & 133 & -10 & -4 \\ 52 & 11 & 49 & -10 & 119 & -2 \\ -17 & 16 & 3 & -4 & -2 & 130 \end{vmatrix}_{\mathcal{B}o}$$

Table 3: raw tensor **Co** obtained for the monocrystal (GPa).

From the above measurements (Table 2), we obtained the following raw stiffness tensor **Co** for our superalloy specimen (Table 3). The components of this tensor are given in GPa and in the base attached to the specimen. One can also point out that some components are negative. This is admissible once the eigenvalues of this tensor are positives which is true in this case. As said, these matrices (written under Voigt's convention) do not reveal the possible nature of the symmetry of **Co**; this tensor looks triclinic. As expressed in Bo, **Co** cannot, at this step, be compared to the stiffness tensor measured by the micro-hardness method (Table 4) expressed in the natural symmetry base Bs related to the lattice.

$$\begin{vmatrix} 213\,(180) & 149\,(103) & 149\,(103) & 0 & 0 & 0 \\ 149\,(103) & 213\,(180) & 149\,(103) & 0 & 0 & 0 \\ 149\,(103) & 149\,(103) & 213\,(180) & 0 & 0 & 0 \\ 0 & 0 & 0 & 140\,(100) & 0 & 0 \\ 0 & 0 & 0 & 0 & 140\,(100) & 0 \\ 0 & 0 & 0 & 0 & 0 & 140\,(100) \end{vmatrix}_{Bs}$$

Table 4: nearest cubic tensor **Cs** for the monocrystal (GPa). Values in italic are obtained from microhardness tests.

# 3. Visualization of the quasi-symmetries: pole figures

## 3.1. principle

Let **Co** be a "raw" stiffness tensor determined by any experimental method, *e.g.*, the acoustical one presented before. As said in the introduction, this tensor is generally triclinic (without symmetries) due to experimental errors and expressed in the randomly chosen specimen's base Bo. The pole figures classically used in crystallography represent the elements of the symmetry group G. Each symmetry level (of the eight possible (Forte and Vianello, 1996)) has a different set of symmetry planes (Figure 4). Plotting the symmetry planes will allow us to distinguish the symmetry level. Let us introduce the *discrepancy function* d(**Co**, **r**):

$$d(\mathbf{Co}, \mathbf{r}) = \frac{\|\mathbf{Co} - S[\mathbf{r}^\perp](\mathbf{Co})\|}{\|\mathbf{Co}\|} \quad (17)$$

In this expression $S[\mathbf{r}^\perp](\mathbf{Co})$ represents the stiffness tensor, symmetrical to **Co** with respect to the plane $\mathbf{r}^\perp$, orthogonal to **r** (see eqn 17a,b in the appendix)[2]. The norm is taken as the Euclidian one (see eqn 17c in the appendix). The distance d is, of course, null if $\mathbf{r}^\perp$ is a symmetry plane for **Co**. As said, **Co** is here generally triclinic but may be "close" to belonging to a higher symmetry group. In this case, some directions **r** will be such that d(**Co**,**r**) has a "low" value. Let us now plot d(**Co**,**r**) for each **r** of the half space; we obtain the *pole figure of symmetry planes*. We can see (Figures 5 and 6) some of these pole figures. The $x_{o3}$ axis is the normal to sheet plane, $x_{o1}$ is on the right hand, and $x_{o2}$ is upwards. To each "spot" corresponds a plane of "quasi-symmetry". We recall here, for an easier interpretation, the shape of pole figures in the case of the following quasi-symmetries (although this can be read on Figure 4):

---

[2] Remark: the computation is about 10 time faster while using the Bond's matrices [1]

- triclinic: no spots
- monoclinic: one spot
- trigonal: three spots (coplanar, at 120°)
- transverse isotropic: one band (infinity of coplanar normals) and one spot (orthogonal)
- orthotropic: three spots (the normals are orthogonal each other)
- tetragonal: five spots (four coplanar normals at 45° and one orthogonal)
- cubic: nine spots (normals of nine cube's symmetry planes)
- isotropic: all the map is black

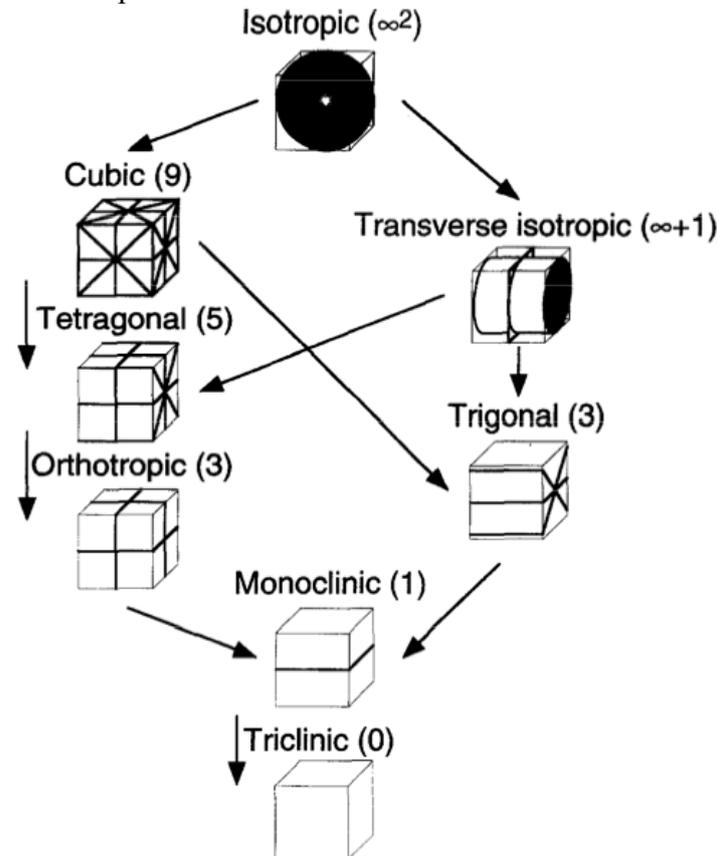

Figure 4: symmetry planes for each symmetry level.

## 3.2. Examples

A brief analysis of Figure 5 reveals the cubic nature of our monocrystal. We can clearly distinguish the nine spots. The cubic symmetry is however non exact, as the value 0 is nowhere reached for d(**Co**, **r**). This cubic symmetry is that of the crystal's atomic structure and we can remark that the $x_{03}$ axis of the specimen is 6 degrees apart from the symmetry axis of the crystal [3].

When the measurements provide less accuracy, the pole figure is not very contrasted and as a consequence different levels of symmetry are possible. We can see that on the pole figure obained from a stifness tensor **Co** measured on an oak specimen (Figure 6). At a first sight transverse isotropy appears (horizontal zone). A closer examination also shows two spots in this zone that reveal a tendency to orthotropy. The ultrasonic measurements are less accurate than those of the monocrystal as the material has some important heterogeneity. The minimum value of d(**Co**,**r**) can be an indicator of the measurements' precision if the material is assumed to have some symmetry.

---
[3] $x_{03}$ is the axis of the cylinder (before machining the specimen), but $x_{01}$ and $x_{02}$ are randomly chosen.

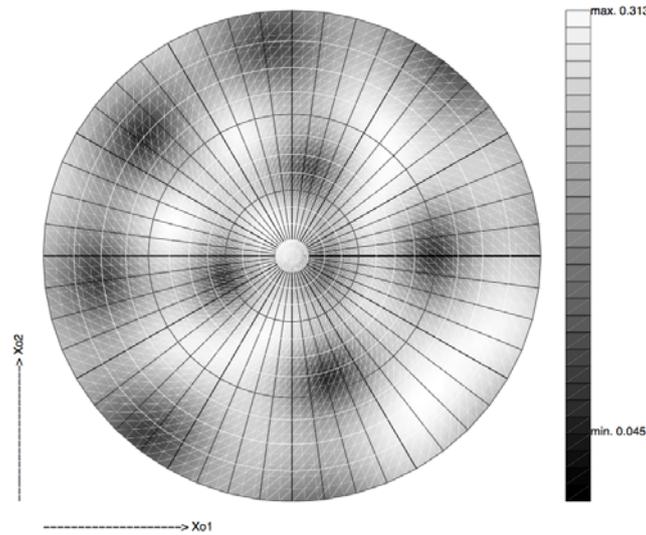
Figure 5: pole figure of symmetry planes obtained from the superalloy's raw stifness tensor **Co**

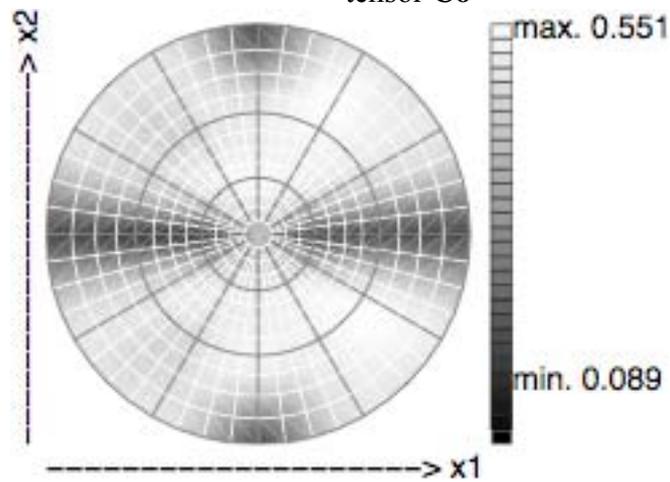
Figure 6: pole figure of symmetry planes obtained from the oak's raw stifness tensor

# 4. Computation of the nearest symmetric stiffness tensor

The last example reveals that a more precise definition of the "distance" between the raw stiffness tensor and the different possible symmetries is necessary for the engineer to choose the most appropriate symmetry level. Furthermore it is necessary to compute the nearest (to **Co**) stiffness tensor **Cs** that belongs to this class of symmetry and its associated natural base Bs (in which **Cs** has the classic matricial Voigt's expression).

## 4.1. general principle

The chosen symmetry for the studied material has a symmetry group G. This group is a subgroup of the orthogonal group O(3) but, as all the (fourth-rank) stiffness tensors have the punctual symmetry -I, one can consider G as a subgroup of SO(3) without loss of information. Most of the notations in the present paper are the ones used by Forte & Vianello (1996).

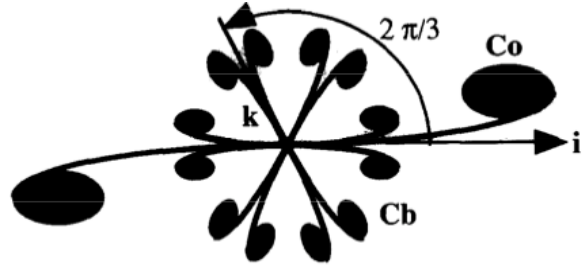

Figure 7: representation of the construction of a trigonal tensor **Cb** from a triclinic tensor **Co**.

Let $G_B$ be a symmetry group G located by an orthonormed base B (in comparison with the specimen's base Bo).
The raw stiffness tensor **Co** is *a priori* triclinic. We now define the *orbit* $G_B$ o **Co** as the collection of all the transformations of **Co** by the elements of $G_B$. The average $<G_B$ o **Co**$>$ of this orbit has, of course, $G_B$ as symmetry group. Let us call **Cb** such an average[4], which is the nearest to **Co** stiffness tensor, having $G_B$ as symmetry group:

$$\mathbf{Cb} = <G_B \text{ o } \mathbf{Co}> \qquad (18)$$

We now can define a distance between **Cb** and **Co**. The following one has the same definition as d(**Co**, **r**) (eqn 17) when G is the symmetry group of the monoclinic symmetry:

$$D(B) = \frac{\|\mathbf{Co} - <G_B \text{ o } \mathbf{Co}>\|}{\|\mathbf{Co}\|} \qquad (19)$$

Then, the distance Ds from **Co** to the considered symmetry group G can be defined as the minimum of the distance D(B) when B varies:

$$Ds = \min_B (D(B)) \qquad (20)$$

This minimum is numerically found by a simplex (Faurre, 1988) method that determines three positioning Eulerian angles. One can remark that this distance does not depend on the arbitrary choice of Bo as the norm above is invariant relative to the choice of the base. The natural base Bs is the argument of the previous minimum:

$$Ds = D(Bs) \qquad (20)$$

And the nearest stiffness tensor **Cs** can now be given as the average of **Co** on its orbit according to $G_{Bs}$:

$$\mathbf{Cs} = <G_{Bs} \text{ o } \mathbf{Co}> \qquad (21)$$

### 4.2. symmetry groups of stiffness tensors
In this section we detail the symmetry group for each symmetry level possible for a stiffness tensor (Forte & Vianello, 1996). As said, we only take into account the elements of SO(3).

**triclinic**
In the case of the triclinic symmetry, the symmetry group G is reduced to the identity {I}. The distance D is of course zero and the base Bs is undefined.

**symmetries based on $D_n$**
We consider as symmetry group $G_B$ the dihedral group $D_n$: in the orthonormed base B=(**i**, **j**, **k**), $D_n$ is generated by $Z_n$ (a cyclic subgroup with n elements, generated by the rotations **Q**(**k**, $2\pi/n$) about **z** of an angle $2\pi/n$) and the rotation **Q**(**i**, $\pi$). $D_n$ has 2n

---
[4] It can be noticed that, as B is positionning the symmetry group $G_B$, **Cb** can generally not be obtained from a rotation of another **Cb'** of this average in comparison to another base B'.

elements. The value of n decides (eqn 18 & 21) of the symmetry class of **Cb** and, of course, **Cs**. We have:
- $n \in \{1, 2\}$ **Cb** is monoclinic
- $n \in \{3, 6\}$ **Cb** is trigonal
- $n \in \{4\}$ **Cb** is orthotropic
- $n \in \{5,7,9,...,\infty\}$ **Cb** is transverse isotropic
- $n \in \{8\}$ **Cb** is tetragonal

In order to clarify this calculation, Figure 7 illustrates how we can obtain a trigonal tensor **Cb** when calculating the average of $D_3 \circ \mathbf{Co}$. The result for the transverse isotropy is justified as a stiffness tensor cannot have these symmetry groups $D_n$ (see theorem 1, Forte & Vianello (1996)). The symmetry group obtained is the transverse isotropy as it contains $D_n$.

**cubic**

The symmetry group $G_B$ of cubic stiffness tensors is O. It is generated, in the base B=(**i**, **j**, **k**), by the dihedral group $D_4$ (defined above) and the rotations **Q**(**i**+**j**+**k**, $2\pi/3$). It has 24 elements.

**isotropic**

In this case the symmetry group $G_B$ of isotropic stiffness tensors is SO(3). As this group has an infinite number of elements, this result cannot give directly the nearest stiffness tensor **Cs**. We propose here two ways to reach the isotropic symmetry.

The first possibility is to observe that O is a maximal subgroup of SO(3). That means that SO(3) is generated by O and every subgroup S not included in O. This method gives an isotropic stiffness tensor **Cs**.

The second method proposed here is to take as $G_B$ the symmetry group of the dodecahedron $I$. It has 60 elements in the following rotations: the rotations **Q**($\mathbf{n}^i$, $2\pi/5$) around the 12 normals of the faces $\mathbf{n}^i$, the rotations **Q**($\mathbf{c}^j$, $2\pi/3$) around the 12 corners $\mathbf{c}^j$ and the rotations **Q**($\mathbf{e}^k$, $2\pi/2$) around the 30 edges $\mathbf{e}^k$. It can be understood that as the stiffness tensor **C** cannot have the dodechedral symmetry (Theorem 1, (Forte and Vianello, 1996)), the following group is SO(3) himself. The result is, of course, independent of the choice of the position of B (*i.e.* the position of the dodecahedron or it's conjugated dodecahedron).

### 4.3. exemple

The stiffness tensor **Co** measured from the superalloy measurements (Table 3) is now considered. The Table 5 gives the distance Ds to each level of symmetry. We can notice that **Co** is really close to the left branch of the tree (the "orthotropic branch"); on the contrary, the "trigonal branch" is far from **Co**. It is obvious that the order relation between the symmetry levels has to be respected. Practically this requires avoiding *local minima* while searching for the base Bs (eqn 20); the position of base B is first manually set close to the real *minima* (using the pole figures of symmetry plane). For low level symmetries it is generally necessary to try many combinations before finding the absolute minima.

|            |        | isotropic           | 34.6 % |                     |        |
|-----------:|-------:|:--------------------|:-------|:--------------------|:-------|
|      cubic | 10.5 % | transverse isotropic | 21.8 % |                     |        |
| tetragonal |  9.9 % |                     |        |                     |        |
| orthotropic |  8.1 % | trigonal            | 21.3 % |                     |        |
|            |        | monoclinic          |  4.2 % |                     |        |
|            |        | triclinic           |   0 %  |                     |        |

Table 5: distance from the raw stiffness tensor **Co** to each symmetry level

Comparing the distance from **Co** to each symmetry level of the left branch leads us to notice that **Co** is only 2.4% further to the cubic symmetry than to the orthotropic symmetry. The high level (only three independant constants) cubic symmetry appears, without any other choice constraints, to be the "right" symmetry level for this material. The nearest cubic tensor **Cs** is given Table 4. The monocrystal has a cubic lattice and X-ray diffraction measurements allows us to validate the obtained symmetry and position of the base B. Furthermore, every **Cs** can be compared to data from mechanical tests made using microhardness tests (Espié, 1996). They allow the measurement of the stiffness tensor especially on small specimens by using load-unload paths. These measurements have been done with the hypothesis that the stiffness tensor has a cubic symmetry related to the corresponding symmetry of the microstructure. The values obtained in the two cases are different but are close to be proportionnal (Table 4). This is probably due to the difficulty of performing mechanical tests on these materials that exhibit some microplasticity even for low strains.

## 5.  Conclusion

We propose in this paper a complete method able to provide the entire stiffness tensor of an unknown elastic material and the full analysis of this tensor. The measurement and the analysis can be used separately as many methods are now able to give the raw stiffness tensor.

Our proposed ultrasonic method is therefore easy and seems reliable for homogeneous materials. The experimental set-up is very simple as a pair of transducers, a card plugged in PC, a software are sufficient to perform the measurements. A possible extension of this work concerns the measurements of induced anisotropy, for example due to damage.

The planes symmetry pole figures are, in our opinion, a straightforward visualisation of the symmetry level of the complex elements that are the stiffness tensors. Their application may be extended to other tensors (different levels, or with different indicial symmetries).

Finally, the presented computation of the nearest stiffness tensor **Cs** gives a powerful, complete and fast analysis of the stiffness tensor. Some interesting mathematical developments will be possible in the case of isotropic symmetry generated by the symmetry group of the dodecahedron. The concept of distance to a symmetry group leads to some surpises: in the case of our oak specimen, one can consider it, without any great loss of information, as transverse isotropic (only 5 constants) instead of the classic orthotropic symmetry level (9 constants) that is commonly used for woods.

# Bibliography

Arts, R. (1993). *A study of general anisotropic elasticity in rocks by wave propagation,*. Univ. P. et M. Curie Paris 6, Paris.

Auld, B. A. (1973). *Acoustic fields and Waves in Solids.* John Wiley & Sons,

Baste, S. and Hosten, B. (1990). Evaluation de la matrice d'élasticité des composites orthotropes par propagation ultrasonore en dehors des plans de symétrie. *Rev. Phys. Appl.* **25**, 161-168.

Boehler, J.P., Demmerle, S. and Koss, S. (1994). A new direct testing machine for anisotropic material. *Exp. Mech.* **34** (1).

Burlish, R. and Stoer, J. (1980). *Introduction to Numerical Analysis.* Springer-Verlag,

Castagnède, B., Jenkins, J. T., Sachse, W. and Baste, B. (1990). Optimal determination of the elastic constants of composite materials from ultrasonic wave-speed measurements. *J. Appl. Phys.* **67** (6), 2753-2761.

Chu, Y.C., Degtyar, D. and Rokhlin, S.I. (1994). On the determination of orthotropic material moduli from ultrasonic velocity data in nonsymmetry planes. *J. Acoustic. Soc. Am.* **95** (6), 3191-3203.

Espié, L. (1996). *Étude expérimentale et modélisation numérique du comportement mécanique de monocristaux de superalliage.* École Nationale Supérieure des Mines de Paris, Paris.

Faurre, P. (1988). *Analyse numérique. Notes d'optimisation.* École Polytechnique & Ellipses, Palaiseau.

Forte, S. and Vianello, M. (1996). Symmetry classes for Elasticity Tensors. *J. of Elast.* **43**, 81-108.

François, M. (1995). *Détermination des symétries matérielles de matériaux anisotropes.* Univ. P. et M. Curie Paris 6, Paris.

Harder, S. (1985). Inversion of Phase Velocity for the Anisotropic Elastic Tensor. *J. Geoph. Res.* **90** (B12), 10,275-10,280.

Hayes, M. (1969). A simple statical approach to the measurement of the elastic constants in anisotropic media. *J. Mat. Sci.* **4**.

Jaric, J. P. (1994). On the condition for the existance of a plane of symmetry for anisotropic material. *Mech. Res. Comm.* **21** (2), 153-174.

Van Bursirk, W. C., Cowin, S.C. and Carter, R. (1986). A theory of acoustic measurement of the elastic constants of a general anisotropic solid. *J. Mat. Sci.* **21**, 2759-2762.

# Appendix: Cartesian notation with indices

If not specified, the convention summation follows on indices. $\delta_{ij}$ is the Kronekker symbol and $\pi_{ijk}$ the direct permutation symbol.

$$\Gamma_{pq} = n_i \, C_{pijq} \, n_j \tag{A1}$$

$$\Gamma_{pq} = \sum_{i=1}^{3} \rho (V^i)^2 (u^i)_p (u^i)_q \tag{A2a}$$

$$\hat{\Gamma}_{pq} = \sum_{i=1}^{3} \rho (\hat{V}^i)^2 (\hat{u}^i)_p (\hat{u}^i)_q \tag{A2b}$$

$$j(\mathbf{n}) = \frac{1}{2} (\Gamma_{pq}(\mathbf{n}) - \hat{\Gamma}_{pq}(\mathbf{n})) (\Gamma_{pq}(\mathbf{n}) - \hat{\Gamma}_{pq}(\mathbf{n})) \tag{A3}$$

$$\frac{\partial j(\mathbf{n})}{\partial C_{ijkl}} = n_i \, (n_p \, C_{pjkq} \, n_q) \, n_l - \sum_{I=1}^{3} \hat{\rho}(\hat{V}^I)^2 \, n_i \, \hat{u}_j^I \, \hat{u}_k^I \, n_l \tag{A4}$$

$$\frac{\partial j(\mathbf{n})}{\partial \varphi_\alpha} = - (\Gamma_{pq}(\mathbf{n}) - \hat{\Gamma}_{pq}(\mathbf{n})) \sum_{I=1}^{3} \hat{\rho}(\hat{V}^I)^2 \left( \hat{u}_p^I \frac{\partial \hat{u}_q^I}{\partial \varphi_\alpha} + \frac{\partial \hat{u}_p^I}{\partial \varphi_\alpha} \hat{u}_q^I \right) \tag{A6}$$

with
$$\frac{\partial \hat{u}_i^\alpha}{\partial \varphi_\beta} = \delta_{\alpha\beta} \, (- \sin\varphi_\alpha \, m_i^\alpha + \cos\varphi_\alpha \, n_i) \tag{A7}$$

and
$$\frac{\partial \hat{u}_i^1}{\partial \varphi_\alpha} = v_i^\alpha - \hat{u}_i^1 \, (v_p^\alpha \, \hat{u}_p^1) \tag{A8}$$

with
$$v_i^\alpha = \frac{\left[ \dfrac{\partial (\pi_{ijk} \, \hat{u}_j^2 \, \hat{u}_k^3)}{\partial \varphi_\alpha} \right]}{\sqrt{\pi_{tpq} \, \hat{u}_p^2 \, \hat{u}_q^3 \, \pi_{trs} \, \hat{u}_r^2 \, \hat{u}_s^3}} \tag{A9}$$

$$L_{ij} = (\hat{u}_p^i \, n_q) \, C_{pqrs} \, (n_r \hat{u}_s^j) - \delta_{ij} \, \hat{\rho}(\hat{V}^i)^2 \tag{A10}$$

$$\frac{\partial h(\mathbf{n})}{\partial C_{pqrs}} = \sum_{i,j}^{\text{known}} L_{ij} \, n_p \, \hat{u}_q^i \, \hat{u}_r^j \, n_s \tag{A13}$$

$$\frac{\partial h(\mathbf{n})}{\partial \varphi_\alpha} = \sum_{i,j}^{\text{known}} L_{ij} \left( C_{pqrs} \left( n_p \frac{\partial \hat{u}_q^i}{\partial \varphi_\alpha} \hat{u}_r^j \, n_s + n_p \, \hat{u}_q^i \frac{\partial \hat{u}_r^j}{\partial \varphi_\alpha} n_s \right) \right) \tag{A14}$$

$$(\mathcal{S}[\mathbf{r}^\perp](\mathbf{Co}))_{ijkl} = S_{ip} \, S_{jq} \, S_{kr} \, S_{ls} \, Co_{pqrs} \tag{A17a}$$

$$S_{ij} = \delta_{ij} - 2 \, r_i \, r_j \tag{A17b}$$

$$\| \mathbf{Co} \| = Co_{ijkl} \, Co_{ijkl} \tag{A17c}$$